\documentclass[prl,showpacs,superscriptaddress,twocolumn]{revtex4}

\usepackage{graphicx}
\usepackage{amsmath}
\usepackage{amssymb}
\usepackage{mathrsfs}

\begin{document}

\title{Anomalous diffusion of phospholipids and cholesterols in a lipid bilayer
and its origins}

\author{Jae-Hyung Jeon}
\affiliation{Department of Physics, Tampere University of Technology,
FI-33101 Tampere, Finland}
\author{Hector Martinez-Seara Monne}
\affiliation{Department of Physics, Tampere University of Technology,
FI-33101 Tampere, Finland}
\author{Matti Javanainen}
\affiliation{Department of Physics, Tampere University of Technology, FI-33101
Tampere, Finland}
\author{Ralf Metzler}
\affiliation{Department of Physics, Tampere University of Technology,
FI-33101 Tampere, Finland}
\affiliation{Institute for Physics \& Astronomy, University of Potsdam,
D-14476 Potsdam-Golm, Germany}

\begin{abstract}
Combining extensive molecular dynamics simulations of lipid bilayer systems of
varying chemical composition with single-trajectory analyses we systematically
elucidate the stochastic nature of the lipid motion. We observe 
subdiffusion over more than four orders of magnitude in time, clearly stretching
into the sub-microsecond domain. The lipid motion delicately depends on the
lipid chemistry, the lipid phase, and especially on the presence of
cholesterol. We demonstrate that fractional Langevin equation motion
universally describes the lipid motion in all phases including the gel phase,
and in the presence of cholesterol. The
results underline the relevance of anomalous diffusion in lipid
bilayers and the strong effects of the membrane composition.
\end{abstract}


\maketitle

Recent advances in single molecule spectroscopy unveil anomalous diffusion of
microscopic tracers in the crowded environment of living cells, starting to
reshape our views of molecular cell biology and underlining the role of
modern statistical physics \cite{pt}. Extensive experimental studies show 
subdiffusion in terms of the non-linear scaling in time of the mean squared
displacement (MSD) \cite{ralf1}
\begin{equation}
\label{msd}
\langle\mathbf{r}^2(t)\rangle\simeq K_{\alpha}t^\alpha\hbox{ with }0<\alpha<1,
\end{equation}
where $\alpha$ is the anomalous diffusion exponent and $K_\alpha$ the
generalized diffusivity of physical dimension $\mathrm{cm}^2/\mathrm{sec}^{
\alpha}$. Subdiffusion (\ref{msd}) was reported for various microscopic
tracers under the densely crowded conditions inside living cells
\cite{golding,weber,jeon,seisenhuber,garini} and in control experiments
\cite{weiss,pan}, as well as for proteins in cell membranes
\cite{weiss1,weigel}. These experiments demonstrate the ubiquitous presence of
subdiffusion of a large variety of tracers and crowded environments over
many orders of magnitude in time, but see \cite{note1}. Subdiffusion alters
significantly
the diffusion control of biochemical reactions, and its effects are therefore
far-reaching for a wide range of molecular cellular processes \cite{minton}.
While subdiffusion \eqref{msd} slows down long-distance diffusional
exchange and may affect surface-bulk exchange \cite{irwin}, it may indeed
be beneficial for local interactions in cells \cite{golding,guigas,leila}.
Depending on the magnitude of the exponent $\alpha$ anomalous diffusion may
effect the localization of objects such as chromosomes or membrane
channels \cite{garini,weigel}, and impact on the formation and dynamics of
membrane domains.

Here we study in detail the diffusive behavior of lipids in bilayer systems
through trajectory analysis from extensive molecular dynamics simulations.
We find that in all investigated bilayers the lipids exhibit subdiffusion up to
a few nanoseconds, before a crossover to either normal diffusion or to
persistent anomalous diffusion with a larger exponent. The observed behavior
depends strongly on the phospholipid chemistry, their mixture with
cholesterol, and the bilayer phase (liquid/gel). Subdiffusion ranges at least
up to several hundreds of nanoseconds in the presence of cholesterols. Our
analysis shows that the lipid motion is consistent with viscoelastic subdiffusion
driven by correlated Gaussian noise in both liquid and gel phases, and thus
provides a \emph{unified\/} physical framework for lipid diffusion in
membranes.

Subdiffusion \eqref{msd} is described by several prominent models based on
different physical mechanisms \cite{ralf1,stas2}. In continuous time random
walks (CTRWs) jumps are separated by random waiting times $\tau$ with
heavy-tailed distributions $\psi(\tau)\sim\tau^{-1-\alpha}$ \cite{scher}.
CTRW motion was identified for microbead motion in reconstituted actin networks
\cite{wong}, lipid granules in cellular cytoplasm \cite{jeon}, and protein
channels in plasma membranes \cite{weigel}. In contrast fractional Brownian
motion (FBM) and fractional Langevin equation (FLE) produce ergodic subdiffusion
\eqref{msd} with long-ranged anti-correlation ($0<\alpha<1$) \cite{lutz}
\begin{equation}
\label{FGN}
\langle\Delta\mathbf{r}(t)\cdot\Delta\mathbf{r}(0)\rangle\sim\alpha(\alpha-1)
t^{\alpha-2}
\end{equation}
of spatial displacements $\Delta\textbf{r}$. FBM is defined by an overdamped
Langevin equation driven by athermal, external Gaussian noise with power-law
correlation. FLE is a generalized
Langevin equation driven by the same noise. Due to its memory kernel the FLE
describes thermal motion of a particle in viscoelastic media \cite{igor}. While
FLE in the overdamped limit produces FBM-like subdiffusive motion, below the
momentum relaxation time, FLE motion is ballistic. FBM/FLE motion describe
subdiffusion in living cells of mRNA \cite{magdziarz}, chromosomal loci
\cite{weber}, and of lipid granules at longer times \cite{jeon}, as well as
the motion of macromolecules in a crowded dextran solution \cite{weiss}. While
sharing the scaling form of the MSD \eqref{msd}, CTRW and FBM/FLE lead to
completely different dynamics of diffusion-control. Knowledge of the
subdiffusion mechanism in membranes is therefore vital to advance our
understanding of their physical and biochemical properties.

Lipid bilayers are quasi two-dimensional, highly packed systems made up of
phospholipid molecules, which undergo thermally driven lateral diffusion
and thus constantly reorganize the membrane. The lateral MSD
of membrane lipids typically spans three distinct regimes: short-time
ballistic ($\alpha=2$), intermediate subdiffusive ($0<\alpha<1$), and
long-time Brownian motion ($\alpha=1$) \cite{flenner,armstrong}.
The long-time diffusive motion of various kinds of phospholipid
molecules in lipid bilayers has been extensively studied \cite{vaz2,almeida}.
Diffusion of lipids in pure bilayers occurs both in the liquid disordered
and the gel phases below the melting temperature, the latter with decreased
diffusivity. Moreover, in bilayers mixed with cholesterols, the diffusivity
of the lipids tends to decrease with higher cholesterol
concentration.

Lipid subdiffusion at shorter time scales is comparatively poorly understood.
In the traditional microscopic picture, the lateral movement of lipid molecules
is assumed to occur through jumps when sufficient void space is thermally
activated at nearest sites \cite{almeida,vaz}. Between jumps the molecule,
caged by its neighbors, undergoes rattling motion. This CTRW-type
jump-diffusion model has been used to estimate the diffusivities
of lipids in the liquid-disordered phase and in bilayers
containing cholesterol \cite{vaz2,almeida}. However, 
atomistic simulations \cite{falck} and quasi-elastic neutron scattering
experiment \cite{busch} showed that such jump-like
displacements rarely occur, and the lipids move concertedly with their
neighbors as loosely defined clusters. Moreover, conflicting
results were reported on the stochastic nature of the lipid diffusion:
Refs.~\cite{flenner,kneller} demonstrated that the lipid motion is consistent
with FLE dynamics, whereas Ref.~\cite{akimoto} claimed to observe CTRW-type
motion governed by non-Gaussian fluctuations and scale-free rattling dynamics.

\begin{figure}
\includegraphics[width=8.8cm]{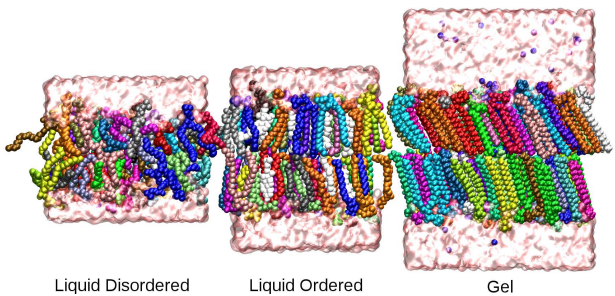}
\caption{Final configurations of simulations of DOPC 338 K (Left),
DSPC/cholesterols 338 K (Middle), and DSPC 310 K (Right) in the liquid
disordered, liquid ordered, and gel phases, respectively (note the
difference in packing states). Each color represents a different
phospholipid. Explicit water molecules correspond to the upper and lower
transparent coatings. Cholesterols appear in white (Middle), Na$^+$ and
Cl$^-$ ions as blue spheres (Right).
\label{snapshots}}
\end{figure}

Lipid bilayers of 128 phospholipid molecules were studied by molecular dynamics
simulations under periodic boundary conditions, for details see the
Supplementary Material (SM) \cite{supp}. We used three pure single component
lipid bilayers composed of DSPC, SOPC, and DOPC phospholipids in the liquid
disordered phase \cite{REM}. We also studied these systems with additional 32
cholesterols (20\% molar concentration) in the liquid ordered phase.
A pure membrane of 288 DSPC molecules was also simulated in the gel phase.
Fig.~\ref{snapshots} shows typical snapshots in the three phases. In this work
we focus on the characterization of the lipid diffusion. To that end we note
that during the
simulation the center of mass of the upper and lower lipid layers undergo free,
independent translational motion (Fig.~S1), as reported previously
\cite{akimoto,edholm}. Free center of mass diffusion causes apparent normal
diffusion of individual lipid molecules at longer times, irrespective
of their actual diffusion characteristics. To avoid this we analyze the relative
motion $\mathbf{r}(t)$ from the center of mass of lipids and cholesterols. 
Figs.~S2, S5, and S11 in SM show sample trajectories.

\begin{figure}
\includegraphics[width=8.8cm]{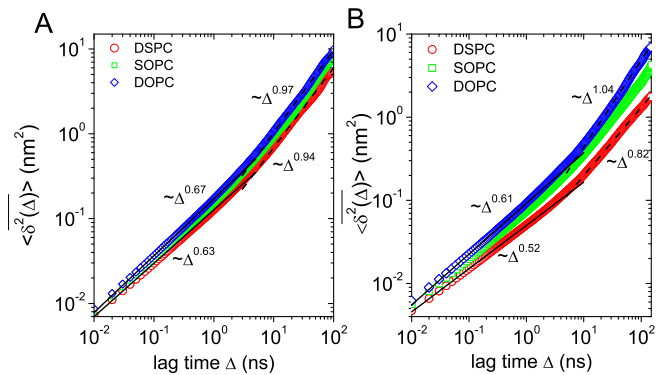}
\caption{TA MSDs $\langle\overline{\delta^2(\Delta)}\rangle$ of DSPC, SOPC, and
DOPC in liquid phase bilayers. Each curve represents the mean of individual
$\overline{\delta^2(\Delta)}$ taken over all trajectories of 128 phospholipids
in the bilayer. A. Cholesterol-free case. B. With cholesterol. The results
were fitted separately to $\langle\overline{\delta^2(\Delta)}\rangle=4K_\alpha
\Delta^\alpha$ in the regimes of short [0.01ns$\ldots$1ns] and long times
[10ns$\ldots$100ns].  Fit results for $\alpha$ and $K_{\alpha}$ are
summarized in Tab.~\ref{table1}.
\label{eatamsd}}
\end{figure}

From individual trajectories $\mathbf{r}(t)$ we obtained the time-averaged
(TA) MSD of lipids typically defined as \cite{golding,weber,stas2}
\begin{equation}
\overline{\delta^2(\Delta)}=\frac{1}{T-\Delta}\int_0^{T-\Delta}[\mathbf{r}(t+
\Delta)-\mathbf{r}(t)]^2dt,
\label{tamsd}
\end{equation} 
where $\Delta$ is the lag time and $T$ the length of the trajectory (measurement
time). Fig.~\ref{eatamsd} shows the mean $\langle\overline{\delta^2(\Delta)}
\rangle$ taken over the trajectories of all phospholipids, for the cases of
DSPC, SOPC, and DOPC in absence and presence of cholesterol. In each case, the
result was fitted by $\langle\overline{\delta^2(\Delta)}\rangle=4K_\alpha\Delta^
{\alpha}$ at short and long times, respectively. The corresponding diffusion
exponents $\alpha$ and diffusivities $K_\alpha$ are summarized in
Tab.~\ref{table1}. In Fig.~\ref{eatamsd}, the scaling behaviors for pure
DSPC and DOPC at short (solid line) and long (dashed line)
times are indicated. In absence of cholesterol, all three
types of lipid molecules show similar behavior: anomalous diffusion
with exponent $\alpha\sim0.6$ below a crossover time $\tau_c\sim10$ ns,
and normal Brownian motion beyond $\tau_c$. The crossover time $\tau_c$
roughly corresponds to the diffusion time of a lipid molecule needed to span
its nearest-neighbor distance. The structural difference in the tails of the
lipids affects somewhat the long-time diffusion, in particular, the values of
$K_{\alpha_l}$.

\begin{table}
\begin{center}
\begin{tabular}{c|c|c|c|c}
\hline
{} & $\alpha_s$ & $K_{\alpha_s}$  & $\alpha_l$ &
$K_{\alpha_l}$ \\
\hline
\hline
DSPC & 0.63(0.06\%) & 0.032(2\%) & 0.94(0.9\%) & 0.020(5\%)\\\hline
SOPC & 0.66(1\%)& 0.038(2\%) & 1.00(16\%) & 0.020(60\%)\\\hline
DOPC & 0.67(2\%)& 0.043(2\%) & 0.97(13\%) & 0.028(16\%)\\\hline\hline
\textbf{DSPC} & 0.52(1\%) & 0.013(2\%) & 0.82(4\%) & 0.0076(6\%)\\\hline
\textbf{SOPC} & 0.58(0.4\%) & 0.019(0.5\%) & 0.87(5\%) & 0.012(4\%) \\\hline
\textbf{DOPC} & 0.61(0.5\%) & 0.023(0.4\%) & 1.04(22\%) & 0.0098(41\%)\\\hline
\end{tabular}
\end{center}
\caption{Exponent $\alpha$ and diffusivity $K_\alpha$ (nm$^2$/ns$^{\alpha}$)
of lipids at short ($s$) and long ($l$) times. Statistical uncertainty in
parenthesis. Boldface: systems containing 20\% cholesterol.
\label{table1}}
\end{table}

Fig.~\ref{eatamsd}B shows that cholesterol significantly affects both the short
and long time diffusion of the lipids. Especially for saturated DSPC with the
smallest cross section area \cite{hector2}, we observe that below $\tau_c
\approx$ 10 ns $\alpha$ decreases to about 0.5 and a new subdiffusion regime
emerges with $\alpha\sim0.8$ up to 100 ns. An additional 1 $\mu$s-long simulation
confirms that this new regime above $\tau_c$ is in fact a slow transition toward
normal diffusion that lasts over hundreds of nanoseconds (Fig.~S10). In the
displayed time window the behavior is well fitted by the scaling exponents
$\alpha_l$ listed in Tab.~\ref{table1}. For unsaturated DOPC, the effect of
cholesterols is small, albeit $K_{\alpha}$ is significantly reduced, and no
second subdiffusion regime occurs.

The observed subdiffusive behavior of lipids is mainly attributed to the unique
structural complexity of the lipid molecule. Spherical-shaped hard particle
systems cannot have such a long subdiffusion regime and values of $\alpha$
as small as 0.5-0.6 (Fig.~S18). To gain additional physical insight into the
lipid motion we now check the detailed stochastic properties of the lipids.

\begin{figure}
\includegraphics[width=12.4cm]{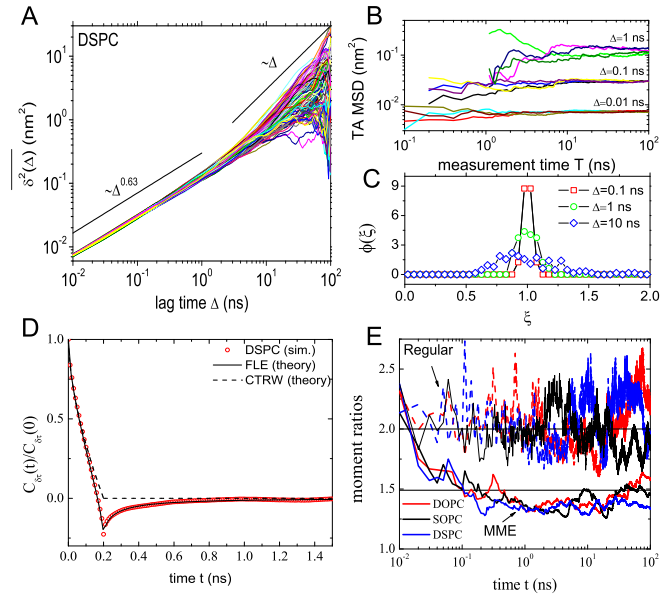}
\caption{Stochastic analysis of DSPC in pure liquid phase bilayer. A. TA MSD
$\overline{\delta^2(\Delta)}$ of all 128 DSPC molecules. B. $\overline{\delta^2(
\Delta)}$ versus measurement time $T$, for $\Delta=0.01$, 0.1, 1 ns. C.
Normalized scatter distribution $\phi(\xi)$ of $\overline{\delta^2(\Delta)}$
versus $\xi=\overline{\delta^2(\Delta)}/\langle\overline{\delta^2(\Delta)}
\rangle$, for $\Delta$=0.1 ns, 1 ns, 10 ns. D. Displacement autocorrelation
function $C_{\delta t}(t)/C_{\delta t}(0)$ of DSPC lipids, $\delta t=0.2$ ns.
The solid and dotted lines represent the fit-free forms of $C_{\delta
t}(t)/C_{\delta t}(0)$ for FLE and CTRW (SM \cite{supp}) at $\alpha=0.63$. E.
Moment ratios $\langle \mathbf{r}^4(t)\rangle/\langle\mathbf{r}^2(t)\rangle^2$
(regular) and $\langle r_{\mathrm{max}}^4(t)\rangle/\langle r_{\mathrm{max}}^2
(t)\rangle^2$ [mean maximal excursion (MME)] for DSPC, SOPC, and DOPC molecules.
The horizontal line at 1.49 distinguishes FLE motion ($\langle r_{\mathrm{max}}
^4(t)\rangle/\langle r_{\mathrm{max}}^2(t)\rangle^2<1.49$) from CTRW
($\langle r_{\mathrm{max}}^4(t)\rangle/\langle r_{\mathrm{max}}^2(t)
\rangle^2>1.49$). The horizontal line at 2 is the expected value of the regular
moment ratio for both FLE and CTRW motion.
\label{dspc}}
\end{figure}

Time-averaged observables obtained from single trajectories provide information
on the ergodic properties of a stochastic motion, and thus about the physical
nature of the underlying dynamics. We call a process ergodic when the long time
average of a quantity (e.g., the MSD) equals the corresponding ensemble average
\cite{stas2,stas1}. For free CTRW subdiffusion \cite{free} $\langle\overline{
\delta^2(\Delta)}\rangle$ grows like $\Delta$, while the ensemble average
\eqref{msd} scales sub-linearly \cite{He,stas2}. Free FLE motion is ergodic
\cite{deng,jae}, and $\langle\overline{\delta^2(\Delta)}\rangle\simeq\Delta^{
\alpha}$. The observation of a sublinear slope in Fig.~\ref{eatamsd} already
indicates that the observed motion is not of CTRW type. This is further
confirmed by the independence of $\overline{\delta^2}$ of the measurement time
$T$, Fig.~\ref{dspc}B, in contrast to the $T^{\alpha-1}$ scaling of CTRW
\cite{stas2,He}. Fig.~\ref{dspc}C shows the distribution $\phi$ of
trajectory-to-trajectory amplitude variations of $\overline{\delta^2(\Delta
)}$ for the 128 individual molecules of Fig.~\ref{dspc}A, as function of the
dimensionless variable $\xi=\overline{\delta^2(\Delta)}/\langle\overline{
\delta^2(\Delta)}\rangle$. All curves are centered around the ergodic value
$\xi=1$. The broadening of $\phi$ with increasing $\Delta$ mirrors large
fluctuations of $\overline{\delta^2}$ at long $\Delta$ due to insufficient
statistics when calculating $\overline{\delta^2}$. $\phi$ also narrows at
fixed $\Delta$ when $T$ is increased (Fig.~S3). These properties demonstrate
that the lipid molecules perform ergodic motion different from CTRW.

\begin{figure}
\includegraphics[width=8.8cm]{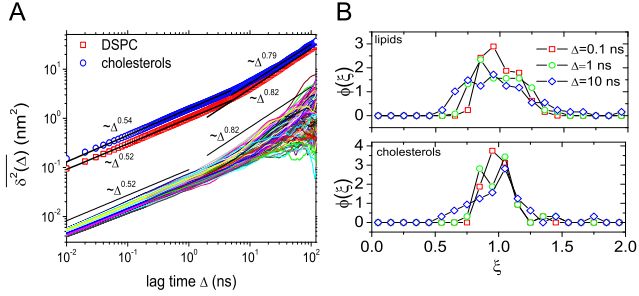}
\caption{DSPC molecules in liquid phase bilayer mixed with cholesterols. A.
$\overline{\delta^2(\Delta)}$ for 128 DSPC molecules (for cholesterols see
Fig.~S8). Average TA MSDs $\langle\overline{\delta^2(\Delta)}\rangle$ for
cholesterols (blue circle) and DSPCs (red square) are shifted by
a factor 20. B. Distribution $\phi(\xi)$ for $\overline{\delta^2}$
of DSPCs and cholesterols.
\label{dspcchol}}
\end{figure}

We obtained the displacement autocorrelation function
\begin{equation}
\label{autocorrelation}
C_{\delta t}(t)=\langle [\mathbf{r}(t+\delta t)-\mathbf{r}(t)]
\cdot[\mathbf{r}(\delta t)-\mathbf{r}(0)]\rangle/\delta t^2
\end{equation}
for arbitrary time step $\delta t$ for several diffusion models in SM III
\cite{supp}. Normalized, $C_{\delta t}(t)/C_{\delta t}(0)$ is a fit free
function for given $\delta t$ and $\alpha$. Fig.~\ref{tamsd}D shows $C_{\delta
t}(t)/C_{\delta t}(0)$ at $\delta t=0.2$ ns of DSPC lipids from simulations in
the subdiffusion regime, along with theoretical results for CTRW and FLE motion
\cite{supp}. We find excellent agreement with FLE motion. Here $\alpha=0.63$ was
taken from the TA MSD. The lipid motion is thus anti-correlated, in line with
Eq.~\eqref{FGN}. The behavior in Fig.~\ref{tamsd}D differs distinctly from 
free CTRW motion \cite{stas2} where $C_{\delta t}(t)=0$ for $t>\delta t$
(dashed line). Fig.~\ref{dspc}E also shows the moment ratios $\langle
\mathbf{r}^4(t)\rangle/\langle \mathbf{r}^2(t)\rangle^2$ and $\langle r_{
\mathrm{max}}^4(t)\rangle/\langle r_{\mathrm{max}}^2(t)\rangle^2$, where
$r_{\mathrm{max}}(t)$ denotes the maximal distance of a given
particle from its initial position reached up to time $t$
\cite{tejedor}. Moment ratios have unique values depending on the
stochastic process \cite{tejedor}, as summarized in SM \cite{supp}.
Fig.~\ref{tamsd}E shows that $\langle\mathbf{r}^4(t)\rangle/\langle\mathbf{r}^2
(t)\rangle^2$ fluctuates around 2, and $\langle r_{\mathrm{max}}^4(t)\rangle/
\langle r_{\mathrm{max}}^2(t)\rangle^2$ decreases from $\approx2$ to stay
$<1.49$, as predicted for FLE and violating CTRW.

We conclude that the subdiffusive behavior shown above is robust, all
analysis tools convincingly point to FLE motion. Analogous results were
obtained for SOPC and DOPC molecules. The results are preserved at varying
temperature: temperature increase only leads to an increase of $K_{\alpha_l}$
in the Brownian regime (Fig.~S4). Consistent with previous studies
\cite{falck,busch} the collective motion of lipids exhibit a flow-like
pattern (Fig.~S16).
   
As shown in Fig.~\ref{eatamsd}, the diffusion of the lipid molecules is
drastically changed by the presence of cholesterols. Is the stochastic nature
also affected by cholesterols? We find that ergodicity of the motion is
preserved, while cholesterols significantly affect the distribution $\phi(\xi)$
of $\overline{\delta^2}$. Comparing with the pure bilayer (Fig.~\ref{dspc}C),
with cholesterol $\phi(\xi)$ noticeably broadens (Fig.~\ref{dspcchol}B).
Individual lipids thus undergo considerable variations while diffusing in the
presence of cholesterols, as seen in Fig.~\ref{dspcchol}A. Moreover, with
cholesterols the displacement autocorrelation $C_{\delta t}(t)$ has a slightly
deeper well, consistent with a stronger anti-correlation of the displacement
(Fig.~S6).

For the motion of the cholesterol molecules themselves we find that their
behavior is almost identical to the lipids', of FLE-type anomalous
diffusion. Fig.~\ref{dspcchol}A compares the TA MSD averaged over all
trajectories $\langle\overline{\delta^2(\Delta)}\rangle$ separately for DSPC
and cholesterol molecules. While cholesterols universally diffuse faster than
the lipids, their scaling behaviors are almost the same. Meanwhile, the
scatter distributions of cholesterols are sharper than that of the lipids,
implying that cholesterol diffusion is more uniform than that of lipids. This
may be related to the fact that at any moment only some lipids are in
direct contact with cholesterols which modifies their behavior from those of
the remaining lipds. The
displacement autocorrelation of cholesterols is hardly different from that
of the lipids in the lipid-cholesterol bilayer (Fig.~S9), and the moment
ratios agree with FLE motion (Fig.~S7).

\begin{figure}
\includegraphics[width=8.8cm]{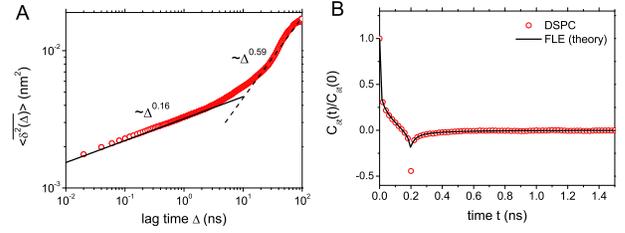}
\caption{DSPC molecules in gel phase bilayer. A. $\langle\overline{\delta^2(
\Delta)}\rangle$. B. Displacement autocorrelation $C_{\delta t}(t)/C_{\delta
t}(0)$ with $\delta t=0.2$ ns. Solid line: theoretical result for FLE
with $\alpha=0.16$.}
\label{dspcgel}
\end{figure}

To obtain a full picture of the diffusive motion in lipid bilayers, we also
studied the gel phase. For DSPC molecules we obtain: (i) Fig.~\ref{dspcgel}A
shows that $\overline{\delta^2}$ scales with $\alpha_s\approx0.16$ at short
times and is thus remarkably smaller than the value $\alpha_s\approx0.6$ in
the liquid phase. Moreover, in the gel phase the TA MSD remains subdiffusive
with $\alpha_l\approx0.59$ beyond the crossover. (ii) The scatter distribution
$\phi$ of individual $\overline{\delta^2(\Delta)}$ shows that the lipid motion
remains ergodic in the gel phase (Fig.~S12) \cite{jae1}. (iii) The results for
the gel phase autocorrelation are consistent with FLE motion with exponent
$\alpha\approx0.16$ (Fig.~\ref{dspcgel}B), as are the moment ratios
(Fig.~S13). (iv) Contrasting recent claims \cite{akimoto}, the rattling
dynamics of lipids is consistent with FLE motion (Figs.~S14, S15).

In summary, we here report extensive molecular dynamics simulations of lipid
bilayer systems and the analysis of individual trajectories using stochastic
analysis tools. While we find a moderate dependence on the lipid chemistry,
the effect of cholesterols is striking. Cholesterols effect more pronounced
and persistent subdiffusion. FLE motion is identified as the
unifying process for the motion of both phospholipids and cholesterols
in liquid and gel phases. Our study thus provides an integral picture of lateral
motion of lipids by showing the compatibility of FLE-type stochastic motion of
individual molecules and their flow-like collective motion.

Cholesterols significantly affect the phospholipids diffusion via increasing
membrane packing and inducing 2D ordering \cite{hector_new}
(Fig.~\ref{snapshots}). $\alpha$ is lowered significantly to $\approx0.5$ below
$\tau_c\approx10$ ns. In agreement, recent experimental and computational
studies show that $\alpha$ decreases with increase of the concentration of
proteins in the bilayer \cite{horton,matti}. Interestingly we observe a
pronounced variation between individual lipid's motion, likely due to the
asymmetric disturbance caused by cholesterols \cite{hector_new}.
While the slowing down of lipid diffusion by cholesterols is known from
experiment \cite{almeida,lindblom} and simulations \cite{edholm},
the dramatic effects of cholesterols on intermediate-time
lipid diffusion have not been reported to our best knowledge.

Given above results we speculate that in biomembranes, whose complexity is
higher than the bilayers' studied here (e.g., larger number of lipid moieties,
proteins, and higher cholesterol concentration), subdiffusion may range to
macroscopic times, thus altering our current view of membrane dynamics. Single
particle tracking together with advanced simulations techniques and stochastic
analysis tools are promising methods to explore this intriguing possibility.

\begin{acknowledgments}
We thank Otto Pulkkinen and Ilpo Vattulainen for discussions. We acknowledge
financial support from the Academy of Finland (FiDiPro scheme) and the CSC--IT
Center for Science in Finland for computing resources.
\end{acknowledgments}


\begin{thebibliography}{99}
\bibitem{free} Free CTRW and FLE motions refer to the processes in free space,
without external confining potential or obstacles that could constrain the
motion.
\end{thebibliography}
\end{document}